# Optomechanical micro-rheology of complex fluids at ultra-high frequency

H. Neshasteh[1], I. Shlesinger[1], M. Ravaro[1], M. Gély[2], G. Jourdan[2], S. Hentz[2], and I. Favero[1*]

[1]Matériaux et Phénomènes Quantiques, Université Paris Cité, CNRS-UMR 7162, 75013 Paris, France

[2]Université Grenoble Alpes, CEA, Leti, F-38000, Grenoble, France

[*] Corresponding author: ivan.favero@u-paris.fr

## Abstract

We present an optomechanical method for locally measuring the rheological properties of complex fluids in the ultra-high frequency range (UHF). A mechanical disk of microscale volume is used as a small-amplitude oscillating probe that monitors the fluid at rest in thermal equilibrium, while the oscillation is detected by optomechanical transduction within a sub-millisecond measurement time, thanks to an optimized signal collection. An original analytical model for fluid-structure interactions is used to extract from these measurements the rheological properties of liquids over the frequency range 100 MHz - 1 GHz. This new micro-rheology method is calibrated by measurements on liquid water, in which we observe pronounced compressibility effects above 500 MHz, but which we show remains Newtonian all over the explored range. In contrast, measurements reveal that liquid 1-decanol exhibits a non-Newtonian behavior, with a frequency-dependent viscosity associated to several relaxation frequencies in the UHF. Our data agree well with an extended Maxwell model for the viscosity of this alcohol, involving two relaxation times of 797 and 151 picoseconds, which are respectively analyzed as supramolecular and intramolecular relaxation processes. A shear elastic response of the liquid appears at the highest frequencies, compatible with the entropic elasticity of molecules as they attempt to uncurl, and whose value enables estimating the volume of a single molecule of liquid. UHF optomechanical micro-rheology demonstrates here a direct mechanical access to the fast molecular dynamics at play in a liquid, in a quantitative manner and in a short time.

The study of the complex dynamics of liquids and viscoelastic materials on various time scales down to picoseconds is of interest for fundamental research and of critical importance for a variety of industrial applications[1–4]. Because commercially available rheometers operate slowly, at frequencies below kHz, the principle of time-temperature superposition has traditionally been used to investigate the fast dynamics of liquids[5–8]. Although the assumption underlying this principle is useful in many cases, particularly in monitoring glass formers, it is not always valid, as evident in many soft materials[6,9–12]. Another approach is to obtain the macroscopic model of the liquid under study, based on molecular dynamics (MD) simulations[13] on a time scale that is not accessible by experimental techniques, for example close to the picosecond range[1,2,14]. Recently, much attention has been paid to this approach, but it reaches computational limits as soon as it is necessary to observe macroscopic dynamics above the nanosecond range, and furthermore, despite the usefulness of MDs, experimental data for verification remain scarce[8,9].

Therefore, to further investigate the fast dynamics of complex liquids and materials, techniques with higher operating frequencies are required. Brillouin scattering, which utilizes energy exchange between photons and thermally driven acoustic waves[15-18], is conveniently employed at frequencies above 5 GHz, but less suited to the UHF range. The frequency shift of Brillouin-scattered photons depends on the acoustic properties of the medium, such as longitudinal modulus and density, but also on its local optical refractive index and thermal properties[17-18]. This leads to an indirect measurement of the local rheological properties. Brillouin scattering is also a weak process, which requires long integration time for analysis. In contrast, rheological measurements based on a solid mechanical body moving in liquid access directly the rheological information of the latter, and can operate with acquisition times well below the second. In consequence, most rheometers have in practice employed small-amplitude vibrating mechanical probes to interact with the liquid. Cantilevers[19,20], micro-electromechanical techniques (MEMS)[21–23], and quartz resonators, have indeed opened access to the rheology of liquid solutions up to a frequency of 100 MHz[9]. However, with these latter electromechanical techniques, it is not possible to further reduce the dimensions of the mechanical probe to achieve higher frequencies. Recently, time-resolved laser pump-probe experiments on vibrating nanoparticles enabled studying fluid-structure interactions between 7 GHz and 140 GHz[24–27]. However, to employ this approach at lower frequencies, the substrate noise must be overcome, which is a formidable task. As of today, a instrumental gap hence exists between 100 MHz and 7

GHz, covering all of the UHF range, where there is no method for analyzing the mechanical motion of a vibrating solid body in a liquid solution (Fig. 1). However many liquids, such as alcohols with long molecular chains, are expected to display a complex dynamics in this frequency range[28], and the associated timescales would additionally be accessible to MD simulations, offering a platform to compare experiments and numerical simulations. The UHF range covers numerous mechanical relaxation frequencies in polymer solutions, and has been little explored in biomechanics as well, essentially due to the absence of proper instruments. Measuring locally, and directly, the rheological properties of a liquid or a soft material over the largest possible frequency range, including the UHF, is also of great interest close to a phase transition, where the fluctuation spectrum broadens, and where microscopic spatial domains can form[29]. A method that would fill the present UHF gap in mechanical micro-rheology, and ideally offer fast acquisition, might hence have implications in the physics of glass formers, in the investigation of biological tissues, and in the science of polymers. We introduce and demonstrate such method here.

We employ suspended disk resonators with a thickness of 220 nm and a radius between 3 μm and 12 μm, and make use of multiple mechanical modes of these devices to develop a novel rheometry approach operating all over the UHF (see Supplements), and focused in the present report on frequencies between 100 MHz and 1 GHz. The mechanical motion of the disk is efficiently transduced thanks to a strong optomechanical coupling to its optical whispering gallery modes (WGM), allowing ultra-sensitive and fast optical measurements (see Figs. S1 and S2). A measurement time as short as 70 μs is sufficient to obtain the desired rheological information in our experiments. The large signal-to-noise data accessible with this optomechanical platform, combined with an analytical description of the hydrodynamics of fluids interacting with the moving disk, enable us extracting the mechanical properties of fluids with accuracy and precision. Experimental data and predictions of the model are first compared in the case of water, taken as a reference Newtonian liquid, in order to calibrate this new rheometry method and assess its level of validity and performance. After this calibration, the method is used on 1-decanol, a monohydric alcohol, in order to reveal its non-Newtonian behavior in the UHF range. The frequency response of optomechanical resonators immersed in the alcohol is measured, and the frequency dispersion of the liquid viscosity and elasticity is deduced from these measurements and from the model. The observed viscoelastic response is connected to the molecular dynamics of 1-decanol, illustrating that optomechanical rheometry can shed light on ultra-fast microscopic processes in the liquid.

Our optomechanical disk resonators, with radius a and thickness h, are fabricated from a silicon-on-insulator (SOI) wafer with a 220 nm thick (100) silicon layer over a 1 μm thick silicon oxide buried layer. They are placed in the vicinity of the optical waveguide (Figure 1(b)). The optomechanical transduction method, which utilizes the mechanically-modulated optical WGMs evanescently coupled to the adjacent waveguide, enables access to the in-plane radial vibration mechanical modes of the disk in various environments, including in liquids[30]. The measured spectrum of the vibration mode of a disk is shown in Figure 1(c), revealing a large added dissipation and frequency downshift when the device is operated in water instead of air. Such frequency response can be described using a damped harmonic oscillator model including the fluidic force $F_{fluid}$ exerted by the liquid onto the vibrating body:

$$\ddot{x} + \gamma_s \dot{x} + \omega_s^2 x = \frac{1}{m_s}(F_{fluid} + \xi) \tag{1}$$

with $\gamma_s$ and $\omega_s^2 = k_s/m_s$ the damping rate and resonant angular frequency of the bare disk in absence of liquid, $k_s$ the spring constant of the associated vibration, $m_s$ the associated mass, and $\xi$ a random Langevin force. Upon immersion, the natural frequency of vibration becomes $\omega$ and the damping $\gamma$. The fluidic force can be modeled with precision for radial vibrations of a disk, using a perturbative approach in which the pedestal is neglected and the device is considered as fully surrounded by the liquid. While this assumption may be employed for radial vibrations of high order and frequency covering all the UHF up to 3 GHz, such as those measured and presented in the Supplements (Fig. S3), it is before all particularly well suited for the fundamental radial breathing mode of the disk, whose displacement amplitude is vanishing near the pedestal, as shown in Figure 1(d). In a recent work[31], we analytically obtained the frequency downshift $\Delta\omega = \omega_s - \omega$ and quality factor $Q \simeq \omega/\gamma$ for this radial fundamental breathing mode, upon immersion in the liquid. To that purpose, we neglected the damping $\gamma_s$ with respect to that exerted by the fluid, and expressed the susceptibility of the immersed resonator by writing Eq. 1, in absence of Langevin force, in the case of a harmonic oscillation $x = Re(\tilde{x} e^{j\omega t})$:

$$(-m_r \omega^2 + j\gamma\omega + \omega_s^2) = 0 \tag{2}$$

where the fluidic force exerted by a compressible viscous liquid resulted in an effective mass $m=m_r m_S$, with $m_r$ a relative mass, and a damping rate $\gamma$ [31]:

$$m_r = 1 + p_1 \frac{\sqrt{\frac{2\mu\rho}{\omega_s}}}{\rho_s h} + p_2 \frac{2\sqrt{\frac{2\mu\rho}{\omega_s}} - \frac{\rho h}{2}\int_0^{2ka} H_0(x)dx}{a\rho_s\left(1 - \frac{J_0(k_s a)J_2(k_s a)}{J_1^2(k_s a)}\right)} + p_3 \frac{\frac{\rho h}{\pi}\ln\left(\frac{32a}{h}\right)}{a\rho_s\left(1 - \frac{J_0(k_s a)J_2(k_s a)}{J_1^2(k_s a)}\right)} \quad (3)$$

$$\gamma = p_1 \frac{\sqrt{2\mu\rho\omega_s}}{\rho_s h} + p_2 \frac{\sqrt{2\mu\rho\omega_s} + \frac{\omega_s \rho h}{2}\int_0^{2ka} J_0(x)dx}{a\rho_s\left(1 - \frac{J_0(k_s a)J_2(k_s a)}{J_1^2(k_s a)}\right)} \quad (4)$$

with the shear viscosity, $\rho$ ($\rho_s$) the density of the liquid (solid), $k=\omega_s/c$ ($k_s=\omega_s/c_s$) the acoustic wave number in the liquid (solid), with $c$ ($c_s$) the speed of sound in the liquid (solid). The longitudinal viscosity, or second viscosity coefficient, can be safely neglected in the UHF range considered here, for reasons discussed in [31]. $J_0(x)$ and $H_0(x)$ are Bessel and Struve functions:

$$J_0(x) = \frac{1}{\pi}\int_0^\pi \cos(x\sin(\varphi))d\phi \quad (5)$$

$$H_0(x) = \frac{1}{\pi}\int_0^\pi \sin(x\sin(\varphi))d\phi \quad (6)$$

The closed-form expressions Eqs. 3 and 4 were derived under certain approximations, and their individual terms correspond to real and imaginary parts of three well-identified contributions in the fluid-structure interactions[31]. We added here an adjustable coefficient $p_{i=1\ldots 3}$ in front of each contribution, in order to precisely fit the experimental data acquired on liquid water, which we consider as a calibration compressible Newtonian liquid over the explored frequency range. The results of this fit are shown in Fig. 2, both for the quality factor and frequency shift, using reference values for water rheological properties (Table 1). A very good agreement is obtained for the calibrated model, which uses the coefficients listed in Table 1. The calibration also led consistent results on water-glycerol mixtures of varying glycerol concentration, hence does not seem to restrict to pure water. In water, if $p_1=p_2=p_3=1$ (raw expression), the uncalibrated model captures the correct trend as function of frequency, but comes numerically off by up to 25%. Note that if compressibility effects in the liquid are neglected (k=0), there is a strong deviation between model and measurements when increasing frequency. Indeed, while dissipation is mainly governed by shear friction for the largest disk radiuses (10<a<12 μm, lowest frequencies), compressibility and acoustic radiation are the dominant mechanisms for smaller radiuses (a<10 μm, higher frequencies): they must absolutely be taken into account by the model.

Table 1: Optomechanical rheological measurements on liquid water

| Coefficients for the calibrated model | | | |
|---|---|---|---|
| | $p_1$ | $p_2$ | $p_3$ |
| | 0.969 | 1.507 | 1.742 |
| Properties of water at 20°c and 1 atm | | | |
| | $\rho$ [kg/m$^3$] | $\mu$ (mPa·s) | c (m/s) |
| Reference values | 1000 | 1 | 1500 |
| OM measurements | 1045±69 | 0.96±0.06 | 1493±58 |

The above frequency-dependent optomechanical (OM) measurements on water allow assessing the accuracy and precision of our method in the evaluation of physical properties of a liquid: density $\rho$, viscosity $\mu$ and sound velocity c. This is achieved by keeping the adjustable coefficients constant, and now minimizing an error function of ($\rho,\mu,$c) when comparing our data with the predictions of the model. The properties of water obtained this way are listed with their uncertainties in Table 1. Uncertainties are calculated from the measurement errors for the quality factor and frequency shift shown in Fig. 2.

With the optomechanical micro-rheology technique now calibrated and characterized on a reference Newtonian liquid, we are prepared to investigate arbitrary liquids at ultra-high frequencies. Long-chain monohydric alcohols are expected to display relaxation dynamics in the UHF range, notably as a result of supramolecular re-organization governed by hydrogen bonds. Recent ultrasonic absorption experiments deduced relaxation frequencies close to the GHz in these alcohols, while the behavior at smaller frequency (up to tens of MHz) was obtained with a fast mechanical rheometer[25]. The frequency stitching of data obtained from the two methods led to partial agreement, with parameters estimated by the two methods varying by a factor up to 5. With our optomechanical rheological method now available, we can instead investigate the response all over the UHF range in a consistent manner. We chose 1-decanol ($C_{10}H_{22}O$) as a candidate for this study.

The measured mechanical response of the optomechanical devices immersed in 1-decanol is shown in Fig. 3. The results are compared with predictions of our analytical model, considering first 1-decanol as a compressible Newtonian fluid with frequency-independent mechanical properties ($\rho$ =827 kg/m$^3$, $\mu$=11.8 mPa s, c=1380 m/s). Fig. 3 clearly shows that such compressible Newtonian assumption fails at describing the measured trends in the UHF range. This is the signature of a viscoelastic response of the liquid, which invites us to explore frequency-dependent laws for the complex viscosity. Having checked that a simple Maxwell model, even involving multiple Maxwell bodies, was not sufficient to reproduce the observed trends, we improved the model complexity in steps, building on concepts of viscoelasticity established at lower frequency. Since the results of Fig. 3 involve at least two inflexion frequencies, a model with at least two characteristic frequencies is required. The simplest viscoelastic model capable to fit our data was found to be a summation of three terms: a first constant shear viscosity $\mu_0$ (Newtonian model), a second term corresponding to the Cole-Cole model[32] with a (slow) relaxation time $\tau$ and a third term associated to an extended Maxwell model[1] of resonant frequency $\omega_0$ and (fast) relaxation rate $\Gamma$:

$$\mu = \mu_\infty + \frac{\mu_0 - \mu_\infty}{1+(j\omega\tau)^n} + \mu_1 \frac{j\omega\Gamma}{\omega_0^2 - \omega^2 + j\omega\Gamma} \tag{7}$$

The Cole-Cole model has been successfully used in polymer science and in geophysics, with values of n between 0 and 2[33,34]. It efficiently grasps physical situations where numerous internal degrees of freedom are involved[33], as we anticipate being the case for the supramolecular dynamics in a long-chain liquid alcohol. As for the extended Maxwell model, it is obtained when taking into account the inertial response of the liquid, as required at the high frequencies considered here (see Supplements and Figs. S3 and S4), and the associated relaxation rate seems to relate to intra-molecular dynamics (see below).

With Eq. 7 for the complex viscosity injected into Eqs. 3 and 4, we obtain a very satisfactory agreement with measurements of the quality factor Q and frequency shift of resonators immersed in 1-decanol (Fig. 3). For that purpose, we inject the complex viscosity into our fluid structure model and minimize an error function with a sequential quadratic optimization algorithm, providing us with an evaluation of the parameters entering Eq. 7. Independent optimization runs provide consistent values for the parameters. The resulting parameters are listed in Table 2, together with parameter variance obtained over 200 optimization runs.

As the here introduced optomechanical rheometry method ventures into the UHF range, literature data for comparison are scarce. In Ref.[35], 1-decanol was studied by two different methods: shear wave impedance spectrometry between 6 MHz and 100 MHz, and acoustic absorption spectroscopy between 300 kHz and 3 GHz. The first method enabled access to the full complex shear viscosity in part of the very high frequency (VHF) range, when the second method attained the UHF range but only led the real part of the viscosity. While not providing the full rheological information in the UHF, the combination of these two methods enabled extracting two relaxation times in the measurement range, on the basis of a fit function with two Maxwell bodies (n=1). By defining in our analysis a relaxation frequency $(\tau^*)^{-1} = \Gamma/2 \left[1 + \sqrt{1 + \frac{4\omega_0^2}{\Gamma^2}}\right]$, where the real viscosity is divided by two, we find $\tau^*$= 151±11 ps, in the range expected for a molecular conformational isomerization (see Supplements). Meanwhile, our value of $\tau$ is consistent with the relaxation of hydrogen-bonded chain-like molecular clusters identified in Ref.[35]. This supports the interpretation of the visco-elastic terms of Eq. 7 as originating respectively from supramolecular (concerning molecule clusters) and intramolecular (within a single molecule) processes, a point that is yet reinforced by the analysis below.

Table 2. Viscoelastic properties of 1-decanol at room T

| Parameters | OM measurements | Ref[35] |
|---|---|---|
| $\mu_0$ (mPa·s) | 14.05±1.31 | 11.8 |
| $\mu_1$ (mPa·s) | 9.24±1.06 | - |
| $\mu_\infty$ (mPa·s) | 3.26±0.85 | 4.9 or ~1 |
| $\tau$ (ps) | 797±38 | 660 |
| $\tau^*$ (ps) | 151±11 | 100 |
| n | 1.25±0.05 | 1 |
| $\omega_0/2\pi$ (MHz) | 625±11 | - |
| $\Gamma/2\pi$ (MHz) | 673±84 | - |
| E (MPa) | 40±8 | - |

Indeed, the introduced optomechanical rheology method enables now a measurement of the full complex viscosity in the UHF range, with its real (µ') and imaginary (µ'') components, providing the complete rheological information. Since $\mu''$ is associated to the elastic response of the liquid, instead of the complex viscosity we rather consider the shear viscosity µ' and shear elastic modulus jωµ''. In Fig. 4 we show the frequency dependence of these two quantities, as obtained from our fit of measurements. At low frequencies, the liquid exhibits a Newtonian behavior with a finite constant viscosity and no shear elastic modulus. At the highest frequencies beyond the GHz, the liquid displays a finite shear elastic modulus E that is plateauing close to 40 MPa. If the high-frequency behavior is governed by intramolecular conformational changes, this limit value of shear elasticity can be assimilated to the entropic elasticity of constituting molecules, which dictates the elongation of a single molecule in its equilibrium conformation[36,37]. Using such argument, the elastic modulus (E) at high frequency can be expressed as inversely proportional to the volume of a single molecule (V):

$$E = \frac{3k_B T}{V} \tag{8}$$

From the inferred shear elastic modulus at high frequency (E=40±8 MPa), the volume of a single molecule of 1-decanol is estimated trough Eq. 8 to be V=2.8 ± 0.7 × $10^{-22}$ cm³. Independently, using the molar mass and density of the liquid, one can determine the occupied volume per molecule. For 1-decanol (molar mass 158.28 g/mol and density 830 kg/m³) the occupied volume is 3.16×$10^{-22}$ cm³, in good agreement with the former. This is evidence that the high-frequency shear elastic modulus of the liquid, and consequently the second relaxation time τ*, are indeed related to the conformational changes and stretching of molecules.

In conclusion, we have developed a new mechanical rheology method operating in the ultrahigh frequency range (UHF). The technique is based on nano-optomechanical resonators and supported by dedicated analytical models of fluid-structure interactions, allowing determination of the liquid properties with very good precision. The approach is used to measure the complex rheological response of a monohydric alcohol, 1-decanol, which shows a non-Newtonian behavior at ultra-high frequency. The exact form of this complex response sheds light on the molecular dynamics at nanosecond timescale in the liquid, accessing both supra- and intramolecular processes. With the UHF frequencies and quality factors reported in the present paper, the response time of our optomechanical rheometer will be sufficient to resolve sub-microsecond time evolutions in real-time. At the same time, our rheological measurements unveil conformational information, leading for example the average volume occupied by a molecule. Because these informations are obtained locally, within a micron-scale measurement volume, the method introduced here is a powerful tool to analyze fast molecular processes in small quantities of liquid and viscoelastic materials. Optomechanical micro-rheology seems notably to possess great assets to track molecular dynamics in the vicinity of the fast phase transition of a micro-droplet, which thanks a small thermal inertia, could be quenched rapidly. This all paves the way for the rheological study of liquid phase transitions, but also for the mechanical investigation of microscale biological objects and membranes in physiological conditions, with unprecedented time-resolution.

**Acknowledgments**

This work was supported by the European Research Council through CoG NOMLI (770933), by the Région Ile de France through the DIM-QuanTiP program.

**Author contribution**

H. N. and I. F. devised experiments and developed the fluid-structure model. H. N and M. R. mounted the set-up. H. N. and I. S. took systematic data. H. N., M. R., I. S. and I. F. analyzed data. G. J, S. H. and I. F. designed the devices. M. G. carried their fabrication. All authors corrected the manuscript, written by H. N and I. F.

**Data availability statement**

The data that support the findings of this study are available from the corresponding author upon reasonable request.

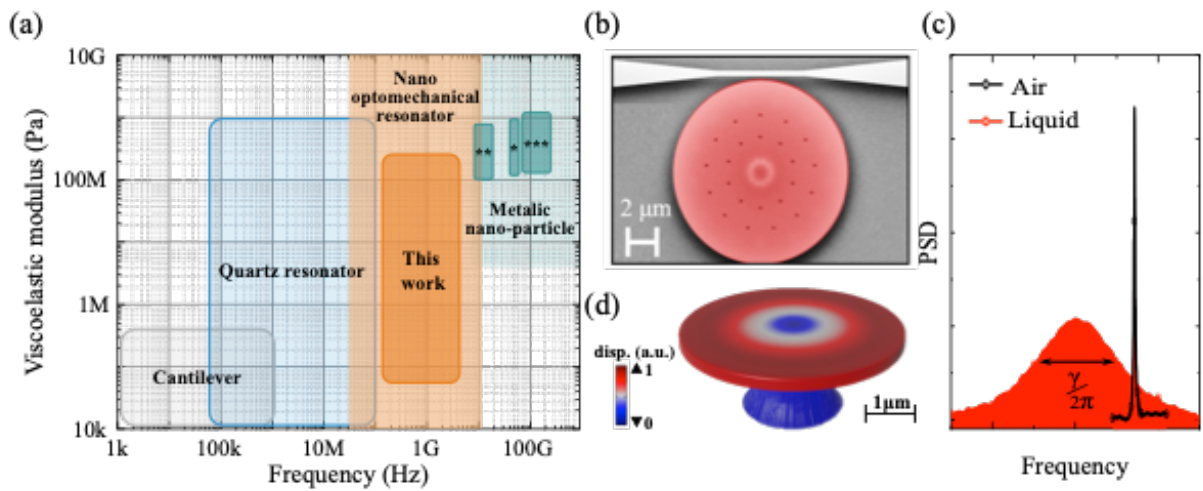

**FIG. 1. Context positioning and technological concept of UHF optomechanical micro-rheology.** (a) Available small-amplitude high-frequency mechanical probes for liquid rheometry. *: ref 26,27**: ref 24, ***: ref 25. (b) Electron micrograph of a suspended miniature silicon disk of radius a=6 μm and thickness h=220 nm, adjacent to a tapered optical waveguide. Release holes have been drilled though the disk to facilitate under-etching of the BOX layer, but play negligible role in relevant fluid-structure interactions. (c) Power spectral density (PSD) of a miniature disk radially vibrating on its fundamental RBM in air (black) and in liquid (red), with a line width of ($\gamma_{air}/2\pi$) and ($\gamma_{liq}/2\pi$), respectively. (d) Distribution of radial displacement for the first RBM of a disk resonator.

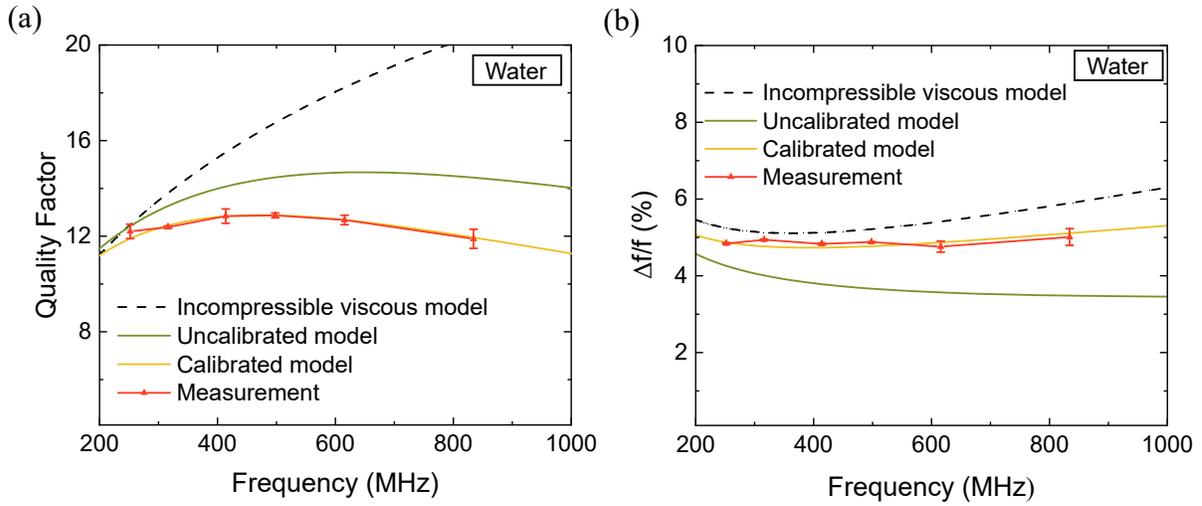

**FIG. 2. UHF optomechanical micro-rheology of liquid water.** Mechanical quality factor and normalized frequency shift in water for the fundamental RBM of a silicon disk of thickness h=220 nm and radius a varying between 3 and 12 µm, corresponding to RBM frequency between 200 and 850 MHz. Measurements in red are compared with predictions of our calibrated (uncalibrated) model in orange (green), and with predictions of an incompressible viscous model in dashed black. (a) Quality factor and (b) frequency shift (decrease) with respect to air.

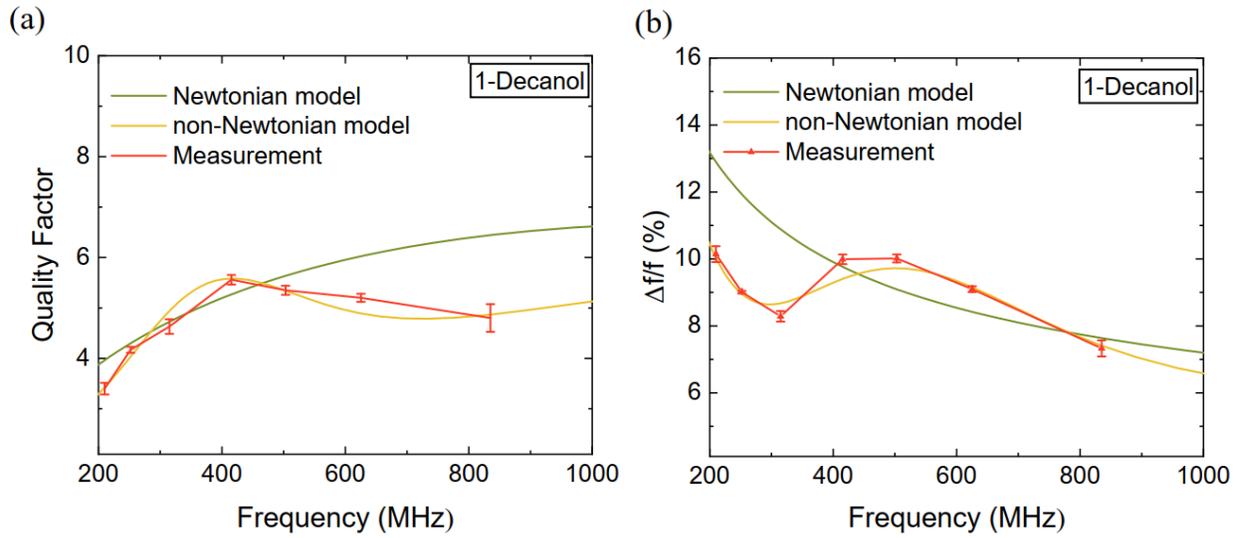

**FIG. 3. UHF optomechanical micro-rheology of liquid 1-decanol.** The mechanical response of miniature silicon disks of thickness h=220 nm and radius varying between 3 µm and 12 µm, once immersed in liquid 1-Decanol. Quality factor (a) and normalized frequency shift (b) as function of frequency. Measured (red) and analytically calculated values, using a Newtonian model (green) and the non-Newtonian model discussed in the text (orange).

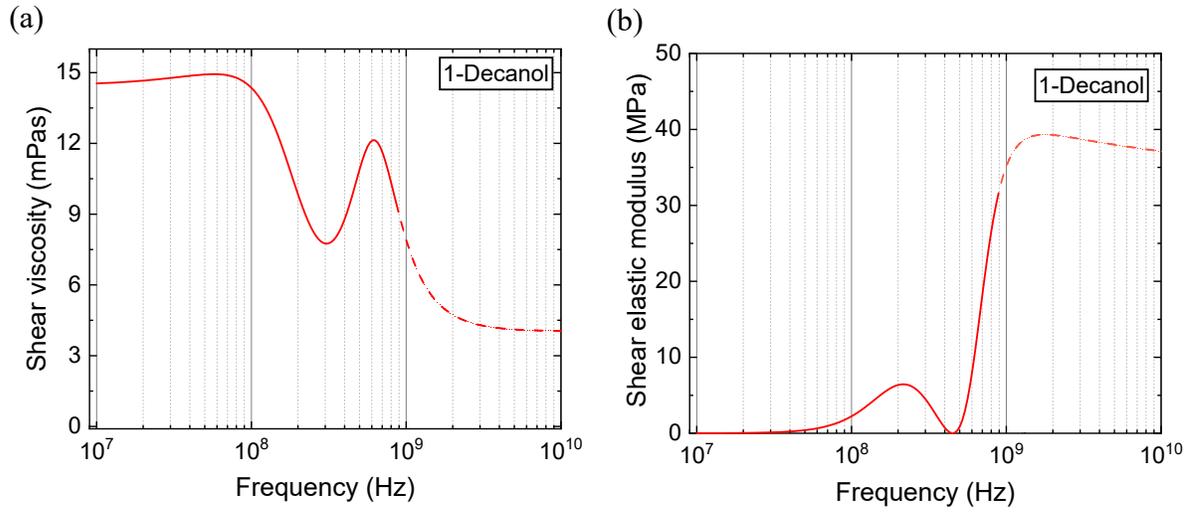

**FIG. 4. The measured complex mechanical response of 1-decanol.** The full rheological behavior of this long-chain alcohol, deduced from our UHF optomechanical measurements and models, is plotted as a function of frequency. (a) Shear viscosity. (b) Shear elastic modulus. The complex viscosity is given by Eq. 7 with parameters obtained from the measurements.

# References

1. Schulz, J. C. F., Schlaich, A., Heyden, M., Netz, R. R. & Kappler, J. Molecular interpretation of the non-Newtonian viscoelastic behavior of liquid water at high frequencies. *Phys. Rev. Fluids* **5**, 103301 (2020).
2. Straube, A. V., Kowalik, B. G., Netz, R. R. & Höfling, F. Rapid onset of molecular friction in liquids bridging between the atomistic and hydrodynamic pictures. *Commun Phys* **3**, 1–11 (2020).
3. Zia, R. N. Active and Passive Microrheology: Theory and Simulation. *Annual Review of Fluid Mechanics* **50**, 371–405 (2018).
4. Deshpande, A. P., Murali, K. J. & Sunil, K. P. B. *Rheology of complex fluids*. (Springer, 2010).
5. Schwarzl, F. & Staverman, A. J. Time-Temperature Dependence of Linear Viscoelastic Behavior. *Journal of Applied Physics* **23**, 838–843 (1952).
6. Olsen, N. B., Christensen, T. & Dyre, J. C. Time-Temperature Superposition in Viscous Liquids. *Physical Review Letters* **86**, 1271 (2001).
7. Van Gurp, M. & Palmen, J. Time-Temperature Superposition For Polymeric Blends. *Rheol. Bull.* (1998).
8. Hecksher, T. *et al.* Toward broadband mechanical spectroscopy. *PNAS* **114**, 8710–8715 (2017).
9. Schroyen, B., Vlassopoulos, D., Van Puyvelde, P. & Vermant, J. Bulk rheometry at high frequencies: a review of experimental approaches. *Rheol Acta* **59**, 1–22 (2020).
10. Peng, X. *et al.* Exploring the validity of time-concentration superposition in glassy colloids: Experiments and simulations. *Phys. Rev. E* **98**, 062602 (2018).

11. Wu, X., Wang, H., Liu, C. & Zhu, Z. Longer-scale segmental dynamics of amorphous poly(ethylene oxide)/poly(vinyl acetate) blends in the softening dispersion. *Soft Matter* **7**, 579–586 (2011).

12. Ding, Y. & Sokolov, A. P. Breakdown of Time−Temperature Superposition Principle and Universality of Chain Dynamics in Polymers. *Macromolecules* **39**, 3322–3326 (2006).

13. Jabbarzadeh, A. & Tanner, R. I. Molecular Dynamics Simulation and Its Application to Nano-Rheology. Rheology Reviews, pp 165-216 (2006).

14. Smith, G. D., Bedrov, D., Li, L. & Byutner, O. A molecular dynamics simulation study of the viscoelastic properties of polymer nanocomposites. *The Journal of Chemical Physics* **117**, 9478–9489 (2002).

15. Magazu, S *et al.*, Relaxation process in deeply supercooled water by Mandelstam-Brillouin scattering, *J. Phys. Chem.* **93**, no. 2, pp. 942–947 (1989).

16. Pinnow, D. A., Candau, S. J., LaMacchia, J. T. & Litovitz, T. A. Brillouin Scattering: Viscoelastic Measurements in Liquids. *The Journal of the Acoustical Society of America* **43**, 1, pp. 131–142 (1968).

17. Comez, L., Masciovecchio, C., Monaco, G., and Fioretto, D. Chapter One - Progress in Liquid and Glass Physics by Brillouin Scattering Spectroscopy. pp. 1–77 in *Solid State Physics*, R. E. Camley and R. L. Stamps, Eds., Academic Press (2012).

18. Carroll, P. J. and Patterson, G. D. Rayleigh–Brillouin spectroscopy of simple viscoelastic liquids. *The Journal of Chemical Physics* **81**, 4, pp. 1666–1675 (1984).

19. Cakmak, O. Precision density and viscosity measurement using two cantilevers with different widths. *Sensors and Actuatorrs A: Physical* **232**, 141 (2015).

20. Ziegler, C. Cantilever-based biosensors. *Anal Bioanal Chem* **379**, 946–959 (2004).

21. Singh, P., Sharma, K., Puchades, I. & Agarwal, P. B. A comprehensive review on MEMS-based viscometers. *Sensors and Actuators A: Physical* **338**, 113456 (2022).

22. Pfusterschmied, G. *et al.* Responsivity and sensitivity of piezoelectric MEMS resonators at higher order modes in liquids. *Sensors and Actuators A: Physical* **295**, 84–92 (2019).

23. Voglhuber-Brunnmaier, T. & Jakoby, B. Electromechanical resonators for sensing fluid density and viscosity—a review. *Meas. Sci. Technol.* 25 (2022).

24. Yu, K. *et al.* Compressible Viscoelastic Liquid Effects Generated by the Breathing Modes of Isolated Metal Nanowires. *Nano Lett.* **15**, 3964–3970 (2015).

25. Yu, K., Yang, Y., Wang, J., Hartland, G. V. & Wang, G. P. Nanoparticle–Fluid Interactions at Ultrahigh Acoustic Vibration Frequencies Studied by Femtosecond Time-Resolved Microscopy. *ACS Nano* **15**, 1833–1840 (2021).

26. Pelton, M., Sader, J. E., Burgin, J., Liu, M., Guyot-Sionnest, P. & Gosztola, D. Damping of acoustic vibrations in gold nanoparticles. *Nature Nanotechnology* *4 (8), 492-495 (2009).*

27. Uthe, B., Collis, J. F., Madadi, M., Sader, J. E. & Pelton, M. Highly spherical nanoparticles probe gigahertz viscoelastic flows of simple liquids without the no-slip condition. *The Journal of Physical Chemistry Letters* (2021).

28. Kaatze, U. & Behrends, R. High-Frequency Shear Viscosity of Low-Viscosity Liquids. *Int J Thermophys* **35**, 2088–2106 (2014).

29. Berthier, L, and Biroli, G. Theoretical perspective on the glass transition and amorphous materials. *Reviews of Modern Physics* **83** (2), 587 (2011).

30. Gil-Santos, E. *et al.* High-frequency nano-optomechanical disk resonators in liquids. *Nature Nanotechnology* **10** (9), 810-816 (2015).

31. H. Neshasteh, M. Ravaro, and I. Favero. Fluid–structure model for disks vibrating at ultra-high frequency in a compressible viscous fluid. *Physics of Fluids* **35,** 5, 059903 (2023).

32. Cole, K. S. & Cole, R. H. Dispersion and Absorption in Dielectrics II. Direct Current Characteristics. *The Journal of Chemical Physics* **10**, 98–105 (1942).

33. Deng. W & Morozov. I. B. Physical Interpretation of the Cole-Cole Model in Viscoelasticity. *Proceedings of the Geoconvention* (2018).

34. Graf, M. J., Su, J.J., Dahal, H. P., Grigorenko, I. & Nussino, Z. The glassy response of double torsion oscillators in solid $^4$He. *Journal of Low Temperature Physics* **162**, 500 (2011).

35. Behrends, R. & Kaatze, U. Hydrogen Bonding and Chain Conformational Isomerization of Alcohols Probed by Ultrasonic Absorption and Shear Impedance Spectrometry. *J. Phys. Chem. A* **105**, 5829–5835 (2001).

36. Tsbolsky, A. V. & Aklonis, J. J. A Molecular Theory for Viscoelastic Behavior of Amorphous Polymers. *The Journal of Physical Chemistry* **68**, 7 (1964).

37. Jones, J. L. & Marques, C. M. Rigid polymer network models. *J. Phys. France* **51**, 1113–1127 (1990).

# Optomechanical micro-rheology of complex fluids at ultra-high frequency: supplements

H. Neshasteh[1], I. Shlesinger[1], M. Ravaro[1], M. Gély[2], G. Jourdan[2], S. Hentz[2], and I. Favero[1]

[1] Matériaux et Phénomènes Quantiques, Université Paris Cité, CNRS, Paris, France

[2] Université Grenoble Alpes, CEA, Leti, Grenoble, France

## Optomechanical measurement principle and set-up details

An optomechanical system typically consists of an optical cavity that embeds a movable mechanical element. An optomechanical disk is the combination of an optical disk supporting whispering gallery modes and of a radial contour disk mechanical resonator. When the disk vibrates on its contour modes (radial breathing for example), the electromagnetic boundary conditions for its optical modes are modulated, producing modulation of the optical response of the system. Fig. S1 illustrates the main ingredients forming a planar optical disk resonator (a), then illustrates the canonical model of an optomechanical Fabry-Pérot resonator (b), and last illustrates the principle of optical detection of mechanical movement in such resonator (c).

In our experiments, a continuous-wave laser at the telecom wavelength is employed to allow optical energy exchange between the optical modes of a bus waveguide and the whispering gallery modes (WGMs) of the disk. One studies the optical response of the disk trough steady-state (DC) laser spectroscopy of the optical cavity, measuring the waveguide optical transmission (output optical signal $P_{out}$) after interaction with the disk, as function of the laser wavelength. By fixing the laser wavelength in contrast, and looking at the high-frequency (AC) modulation of the optical output, one has access to the mechanical response of the disk.

Fig. S2 shows the experimental setup employed to optomechanically measure the properties of liquids. With this set-up we measure simultaneously the DC output optical power and the mechanically modulated AC optical power, hence collecting all the required. The two bottom insets of Fig. S2 display the measured power spectral density (PSD) associated with mechanical vibrations of the disk. To ensure that the fluid remains at rest and does not get affected by the heat generated by the laser, the power absorbed in the silicon disk must be kept as low as possible. To this aim, we keep the input laser power at the lowest level possible, and use of an erbium-doped optical amplifier and an electrical amplifier to amplify the output signal. Even at the highest input powers employed in our experiments, we keep this way the intracavity absorbed power below $100\ \mu W$, which is sufficient to limit such effects.

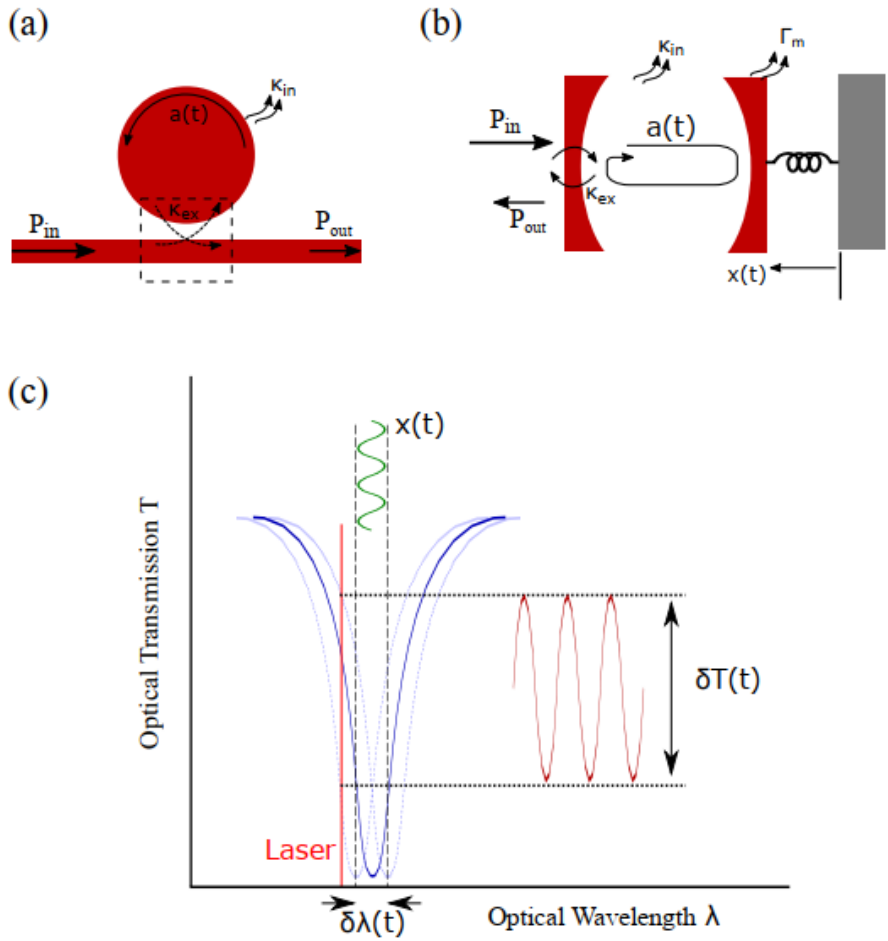

Fig.1. Optomechanical detection of mechanical motion. (a) Standard description of a disk optical resonator coupled to its bus waveguide. $P_{in/out}$ is the incident (output) power. $a(t)$ is the intracavity optical field. $\kappa_{in/ext}$ is the intrinsic (external coupling) decay rate of the optical field. (b) Fabry-Pérot cavity with a movable mirror. The notations are the same as for the disk cavity, but a mechanical resonator is now present, with a position $x(t)$ and a mechanical decay rate $\Gamma_m$. (c) Principle of the optomechanical "on the slope" transduction technique. The laser wavelength (red) is tuned on a flank of the optical resonance (blue). When the position $x(t)$ varies sinusoidally in time, the optical resonance spectrally shifts periodically in consequence (light blue). This produces a periodical variation $\delta T(t)$ of the output optical intensity.

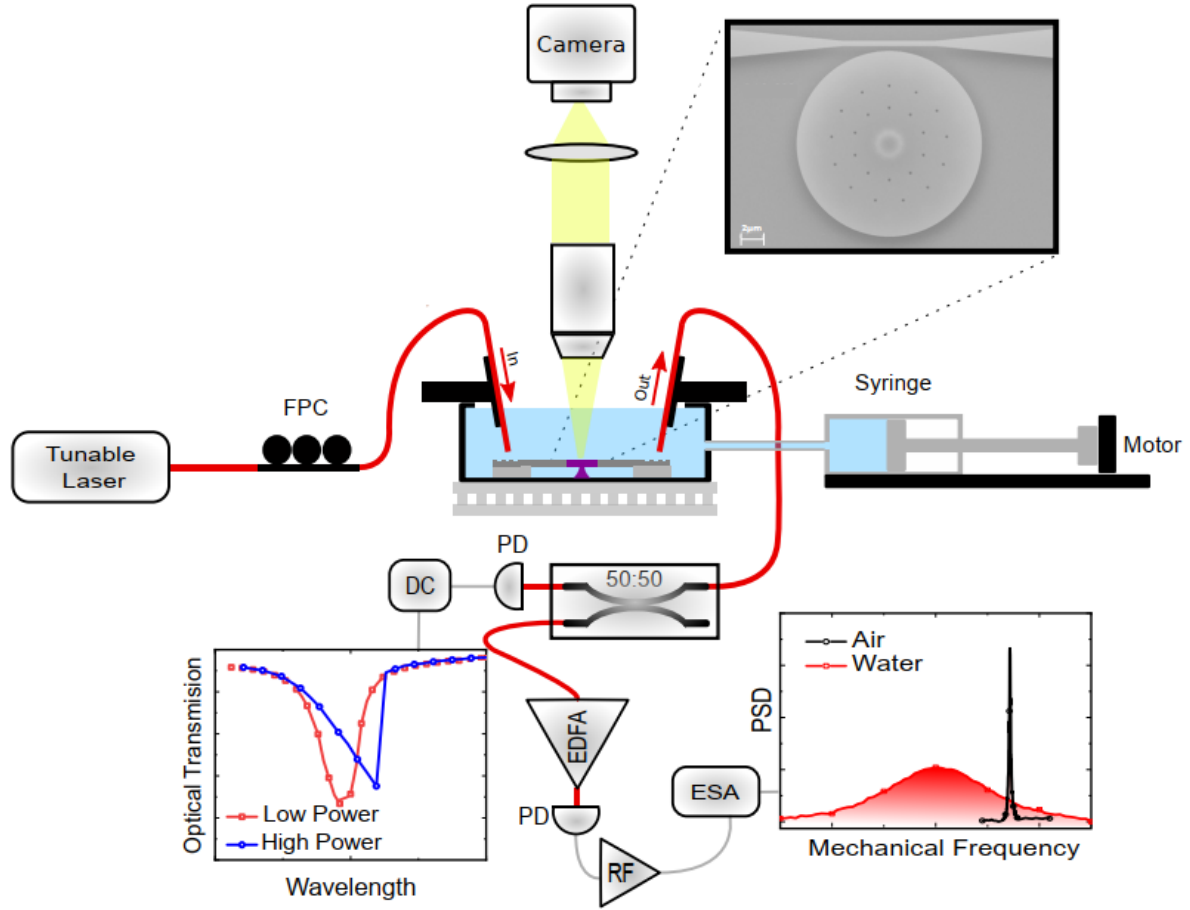

*Fig. S2. Experimental setup to optically measure the rheological properties of liquids within micronic volume. The polarization of the tunable laser pump is adjusted with a fiber polarization controller (FPC). The light is coupled into the integrated waveguide via a tilted fiber above the grating and collected and amplified at the output by an erbium-doped amplifier (EDFA) followed by an RF amplifier behind the photodetector. The DC and the high-frequency (AC) response are obtained on two distinct channels of detection.*

The top inset of Fig. S2 is the top view SEM image of the disk and waveguide. The device is fixed at the bottom of the liquid cell. The light is coupled into the waveguide via a cleaved fiber inclined above the grating structure (schematically shown in Fig. S2). The position of the two input and output fibers is automatically controlled to maintain a constant optical transmission of the waveguide during measurement. After calibrating the sample in air, the cell is filled with liquid using a syringe pump. A top view camera imaging system is used both to determine the best coupling conditions for the fibers and gratings, and to monitor the liquid level. Before measuring the properties of a new liquid, the syringe pump fills first the cell several times with a reference liquid (here DI water is chosen as the reference liquid) and the device response is monitored in order to check the quality of the collected signals. This procedure ensures that the new measurement with a new liquid is meaningful. To monitor and control the temperature of the liquid in the cell, we sue a TEC controller, which is connected to Peltier elements and a thermal resistor on the metallic parts of the liquid cell.

## Measurement of higher-order radial breathing modes

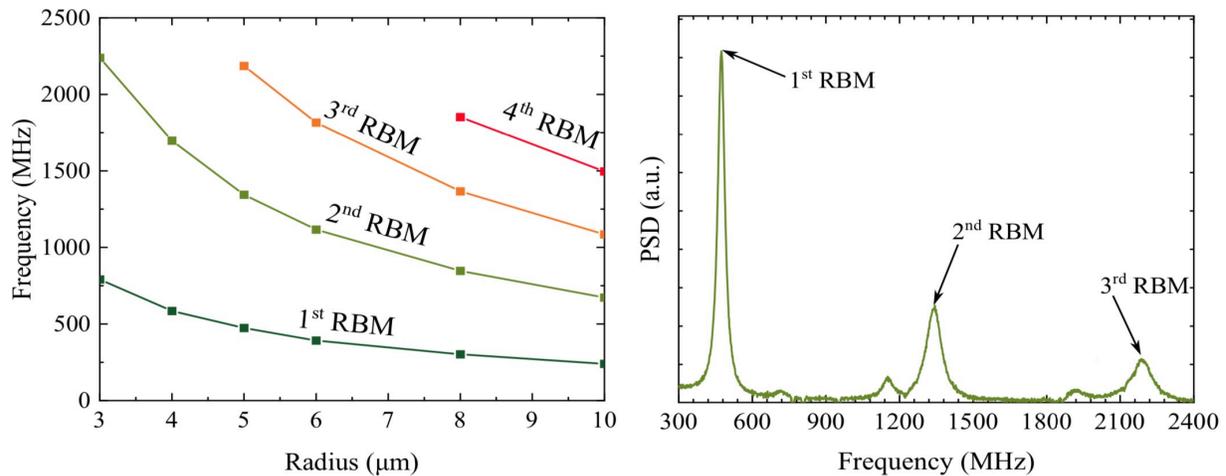

*Fig. S3. Higher-order breathing modes of the disks. <u>Left</u>: Finite Element Method calculation of the Eigenfrequencies of the four first radial breathing modes of a silicon disk of thickness 220 nm, for a radius varying between 3 and 10 µm. The frequencies are evolving between 250 MHz and 2.5 GHz. <u>Right</u>: Spectrum acquired by optomechanical means on a silicon disk of radius 5 µm, enabling observation of the fluid-structure interactions up to 2.2 GHz.*

## Error bars and statistics

There are three sources of error in our data

. Measurement error: We work close to optical resonance, at a certain detuning wavelength, and measure (at least 10 points) at each laser power level, before finally estimating the measurement error.

. Zero optical power estimation error: In order to minimize influence of laser power, we use laser power dependent measurements to extrapolate the response of the device at zero optical power, and we determine the error of this estimation using a curve fitting tool.

. Parameter extraction error: We use our optimization algorithm and run the program many times with different initial conditions to ensure the unicity of the result.

The errors shown in the paper is a convolution of all these errors.

In the study, we used 3 distinct samples (chips), each containing many different resonators of varying radius. In total 30 resonators were measured along this work.

## Derivation of an extended Maxwell model of viscosity

The contribution of intramolecular processes to the linear viscoelasticity of a molecular liquid can be expressed on the basis of the Maxwell model, which links the stress σ to the strain ε (Fig. S4 (a)):

$$\frac{1}{\mu_0}\sigma + \frac{1}{K}\dot{\sigma} = \dot{\varepsilon}$$

This model is based on a viscous damper and a spring in series, but at higher frequencies we should also consider the inertia associated to the mass of the fluid. Considering elements in series, we add a third element relating the strain rate to the time-integral of the stress (Fig. S4 (b)):

$$\sigma + \frac{\mu_0}{K}\dot{\sigma} + \mu_0 L \int \sigma dt = \mu_0 \dot{\varepsilon}$$

With this approach, we can express the complex viscosity (η) as follows:

$$\mu = \frac{1}{\frac{1}{\mu_0} + \frac{j\omega}{K} + \frac{L}{j\omega}}$$

where the conventional Maxwell model is recovered for a fluid with no inertia (L=0). Similarly, in [1], the notion of inertia was used to develop a generalized Maxwell model and recover the non-Newtonian behavior of glycerol-water mixtures predicted by molecular dynamics simulation. Compared to the series circuit model used in our work, in [1] a circuit with a combination of parallel and series components was used instead (see Fig. S4 (c)). The exact combination of circuit elements depends on the details of the molecular dynamics within the liquid.

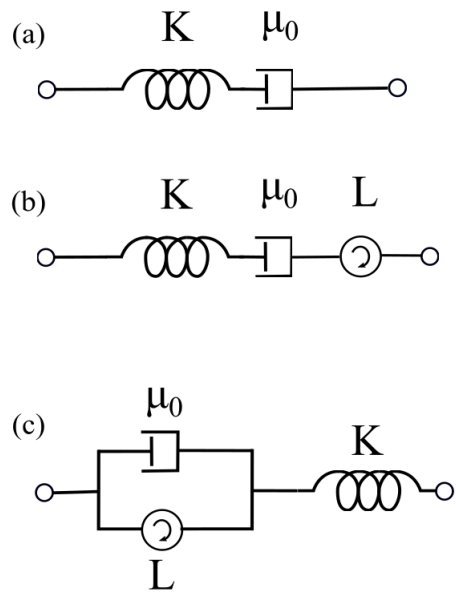

Fig. S4. Circuit representation of linear viscoelastic stress-strain relations. Maxwell model (a) and generalized Maxwell models with an added inertial term: (b) this work, (c) ref [1].

A common treatment for viscoelasticity of polymer solutions consists in representing the polymer by a chain of N subsegments, in the spirit of the Rouse model [2]. In this context, it is usual to consider the overdamped limit for the internal motion of the polymer, from which by assuming a large chain length (N>>1) the relaxation time is found to be [2, 3]:

$$\tau_p = \frac{6\eta_0 M}{\pi^2 p^2 \rho RT}$$

where p is he order of the considered mode of the chain. Trying to describe liquid 1-decanol by such Rouse model, one gets for a single 1-decanol molecule $\tau_{p=1} = 550$ ps. This value is of course a very crude approximation considering the oversimplified derivation and the fact that the Rouse model is not expected to apply to a molecular liquid.

1. Schulz, J. C. F., Schlaich, A., Heyden, M., Netz, R. R. & Kappler, J. Molecular interpretation of the non-Newtonian viscoelastic behavior of liquid water at high frequencies. *Phys. Rev. Fluids* **5**, 103301 (2020).
2. Rouse, P. E. A Theory of the Linear Viscoelastic Properties of Dilute Solutions of Coiling Polymers, *J. Chem. Phys.* **21**, 1272 (1953).
3. Bueche, F. The Viscolelastic Properties of Plastics, *J. Chem. Phys.* **22**, 603 (1954).